\documentclass[aps,pra,twocolumn,groupedaddress,showpacs,amsmath,amssymb]{revtex4}
\usepackage{epsfig}
\usepackage{amssymb}
\usepackage{amsmath}
\usepackage{graphicx}
\usepackage{amsfonts}
\usepackage{epsfig}
\usepackage{revsymb}
\usepackage{bm}
\usepackage{amsmath}
\usepackage{amssymb}
\usepackage{latexsym}
\usepackage{thumbpdf}
\usepackage{hyperref}

\begin{document}

\title{Instability of a crystal $^4$He facet in the field of gravity}

\author{S. N. Burmistrov, L. B. Dubovskii, V. L. Tsymbalenko}
\affiliation{Kurchatov Institute, 123182 Moscow, Russia}


\begin{abstract}
We analyze the analog of the Rayleigh instability  in the field of gravity for the
superfluid-crystal $^4$He interface provided that the heavier $^4$He crystal phase
occupies the half-space over the lighter superfluid phase. The conditions and the
onset of the gravitational instability are different in kind above and below the
roughening transition temperature when the crystal $^4$He surface is in the rough
or in the smooth faceted state, respectively. In the rough state of the surface
the gravitational instability is similar to the classical case of the fluid-fluid
interface. In contrast, in the case of the crystal faceted surface the onset of
the gravitational instability is associated with surmounting some potential
barrier. The potential barrier results from nonzero magnitude of the linear facet
step energy. The size and the tilting angle of the crystal facet are also
important parameters for developing the instability. The initial stage of the
instability can be described as a generation of crystallization waves at the
superfluid-crystal interface. The experiments which may concern the gravitational
instability of the superfluid-crystal $^4$He interface are discussed.

\end{abstract}


\maketitle

\section{Introduction}
\par
The Rayleigh instability is a well-known instability of the interface between two
liquids in the field of gravity when the heavier liquid is placed above the
lighter one. The similar effect is difficult to observe in solids since any
displacement of a piece of the solid body is impeded due to appearing elastic
stresses. However, there exists an elastic stress-free possibility of changing the
shape of a crystal, namely, remelting in the hydrostatic pressure gradient. A
$^4$He crystal in contact with its superfluid phase could be one of most promising
objects to observe the manifestation of the Rayleigh instability for the
liquid-solid interface. In fact, solid $^4$He density is larger than that of the
liquid phase. The growth rate of atomically rough crystal surfaces increases
drastically with the lowering of the temperature, and the change of the crystal
shape can occur in very short time intervals of about 1~s. Unfortunately, in most
of crystal growth experiments \cite{Bal} a $^4$He crystal appears either at the
bottom of an experimental cell or drops on the same bottom later if a crystal seed
nucleates first at the wall. Thus, preparing the necessary configuration with the
heavier crystal above its lighter liquid phase in order to have the initial
condition for the development of the classical Rayleigh instability comes across a
difficulty.
\par
Nevertheless, one experiment, in which the necessary liquid-crystal configuration
against the direction of gravity is arranged, has been reported \cite{De}. A
single hcp $^4$He crystal is grown at the bottom of an experimental cell at the
temperature of about 1.1~K lying between the first and second roughening
transitions. After the grown $^4$He crystal occupies the lower half of the optical
cell, the cell is rotated mechanically through 180$^{\circ}$ so that the $^4$He
crystal proves to be above the superfluid liquid. Then the crystal phase starts to
melt, descending along the cell walls. In its turn, a single finger of the
superfluid phase moves in the upward direction at the center of the cell.
Eventually, the crystal again occupies the lower half of the cell. Unfortunately,
the development of the instability observed is described qualitatively and the
results of visual observations are illustrated by the schematic figure alone. The
main conclusion of the work is that the spatial scale associated with the
development of the instability is about 1~cm.
\par
\begin{figure}
\includegraphics[scale=0.8]{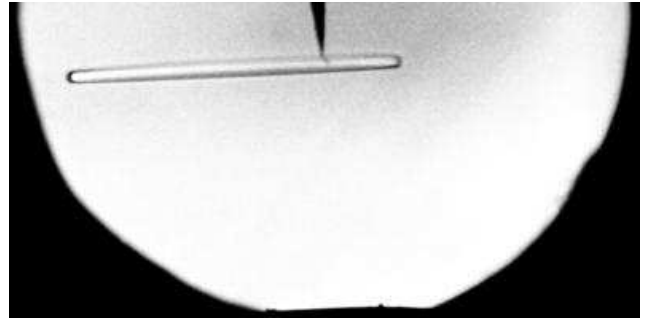}
\caption{The $^4$He crystal grown at 0.92~K between two roughening transitions as
in experiment \cite{Ts}. The diameter of the visible margin is 12~mm.}
\label{fig1}
\end{figure}

This result \cite{De} seems to conflict with the stability observed for the lower
surface of the $^4$He crystals grown on the needle point at the center of a cell
(Fig.~\ref{fig1}).
As is seen from the figure, the lower crystal facet is under conditions
appropriate for the development of the gravitational Rayleigh instability. The
single distinction is likely to be associated with the state of the crystal $^4$He
surface. Whereas in experiments \cite{De} one deals with the rough state of the
crystal $^4$He surface, in our experiments \cite{Ts} one observes the smooth
faceted state of the surface in the form of a hcp crystal basal facet above the
superfluid phase.
\par
In this paper we consider an instability of the crystal $^4$He surface above the
superfluid phase in the field of gravity. We analyze and compare two possible
states of the crystal $^4$He surface such as the rough and smooth faceted ones.

\section{Instability of the crystal surface in the rough state}
\par
We start our consideration from the high temperature region above the roughening
transition when the crystal surface is in the atomically rough state. Since in
this case the surface tension $\alpha$ has no singularity at the close-packed
facets and weakly depends on the crystallographic direction, the tensor of surface
stiffness $\gamma _{ik}$ is close to $\alpha\delta _{ik}$. Thus, to simplify the
analysis, we neglect any distinction between surface tension and surface stiffness
in the formulas for the Laplace pressure.

\subsection{Flat shape of the crystal surface}
\par
Crystallization waves, representing small oscillations of the superfluid-solid
$^4$He interface, are predicted by Andreev and Parshin \cite{An}. The spectrum of
crystallization waves in the field of gravity can be found from the equation
\cite{Ke}
\begin{equation}\label{f1}
\rho _{\text{ef}}\,\frac{\omega ^2}{q}+i\frac{\rho '}{K}\,\omega -\alpha
q^2-\varDelta\rho g =0, \;\;\; \rho _{\text{ef}}=\frac{(\varDelta\rho )^2}{\rho}\,
.
\end{equation}
Here $q$ is the wave vector, $\varDelta\rho =\rho '-\rho$ where $\rho '$ and
$\rho$ are the densities of the solid and liquid phases, $K$ is the interface
growth coefficient, and $g$ is the acceleration of gravity. The positive magnitude
$g>0$ corresponds to the usual situation when the $^4$He crystal lies under the
liquid phase. In this case the frequencies of small interface oscillations have
the negative imaginary parts for all wave vectors and the crystal surface is
always stable.
\par
Provided the solid phase occupies the half-space above the liquid, the dispersion
equation remains the same but parameter $g$ becomes negative. This means that the
last term in~(\ref{f1}), which dominates for sufficiently small $q$, can result in
the positive imaginary part for the roots of the dispersion equation. The critical
magnitude of the wave vector $q_0$ is given by
\begin{equation*}
q_0=\sqrt{\frac{\varDelta\rho\, g}{\alpha}\,}=1/\lambda ,
\end{equation*}
where $\lambda\sim$1~mm is the capillary length. The instability will develop
faster for small wave vectors. For the harmonic $q\rightarrow 0$, the estimate for
the time of developing the instability is given by
\begin{equation}\label{f3}
t\sim\frac{\rho '}{\varDelta\rho}\,\frac{1}{gK}\, .
\end{equation}
For $T\sim$1.1~K, the inverse growth coefficient $1/K\sim$ 2~m/s and $t\sim$ 2~s.
This is in a qualitative agreement with the observations of work~\cite{De}. Thus,
the instability similar to the Rayleigh one develops at the lower surface of a
$^4$He crystal in the atomically rough state.

\subsection{Spherical shape of a solid}
\par
\begin{figure}
\includegraphics[scale=1.0]{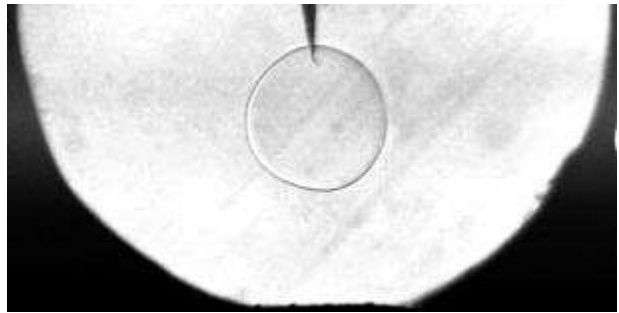}
\caption{The isotropic growth of a $^4$He at 1.28~K. The diameter of the visible
margin is 12~mm.} \label{fig2}
\end{figure}

An interesting situation appears with the nucleation and growth of a crystal at
the needle point at the temperatures above the roughening transitions
(Fig.~\ref{fig2}).
In the hydrostatic pressure gradient  the crystal grows in the shape of a ball
which melts in the upper part and crystallizes in its lower part at the same time.
In outward appearance this looks like the motion of a crystal downward. In work
\cite{Ts}, in which the similar situation is studied, it is shown that the crystal
with the isotropic growth coefficient conserves its spherical shape, and the
descending motion velocity of the sphere $v$ is equal to
\begin{equation}\label{f4}
v=K\,\frac{\varDelta\rho}{\rho}\, gR(t),
\end{equation}
where $R(t)$ is a time-dependent radius of the sphere. Then, as follows from
(\ref{f4}), the center of the crystal shifts downward at a constant velocity in
the hydrostatic pressure gradient provided the crystal volume remains unchanged.
\par
Let us consider stability of the spherical shape of a crystal against small shape
perturbations in the field of gravity. For analysis, we choose the frame which
origin is put at the center of a crystal. In this frame the center of a crystal is
fixed and the liquid phase circulates around the crystal, outflowing from the
upper part of the sphere and flowing into the sphere in its lower part. We
introduce $\zeta (t)\ll R$ as a small perturbation of the radius $R(t)$. The
liquid phase is assumed to be incompressible and the solid one is motionless. In
experiment \cite{Ts} the velocities of the liquid flow are small and do not exceed
0.1~mm/s. For the hydrodynamic pressure $\rho v^2/2$, we have an estimate
10$^{-2}$~ dyne/cm$^2$. This magnitude is much smaller than the typical
hydrostatic pressure drop $\varDelta\rho\, gR$. In what follows, we neglect
quadratic terms in velocity of the liquid phase.
\par
Next, in accordance with the experimental conditions, we take into account that
the experimental cell is closed and no matter comes from outside. In other words,
the total mass of the liquid and solid phases remains unchanged. In the frame
comoving with the crystal the center of the crystal is fixed. As a result, we can
omit the spherical harmonic with $l=0$ from consideration since this harmonic is
associated with the change of the volume. The next harmonic $l=1$ is responsible
for the displacement of the sphere as a whole and corresponds to the circulation
of the liquid around of the sphere. We also omit this harmonic from our
consideration.
\par
The determination of the oscillation spectrum is not difficult but cumbersome. We
give a scheme of solution and then the final result. Let axis $z$ run in the
vertical direction, and we seek for the general solution, expanding perturbation
$\zeta$ in spherical harmonics. Since the liquid is assumed to be incompressible,
it is convenient to describe its motion in terms of velocity potential $\phi$
according to $\bm{v}=\nabla\phi$. The mass flow across the interface is
proportional to the chemical potential difference $\Delta\mu$
 $$
J=\rho 'K\Delta\mu .
 $$
The continuity of the mass flow across the interface allows us to relate the
velocity of the liquid at the interface with the interface growth rate
$\dot{\zeta}$. After some calculations and involving the axial symmetry with
respect to axis $z$, we arrive eventually at the dispersion relation
\begin{multline}\label{f5}
\biggl[\frac{\omega ^2}{\omega _0^2}+i\omega\,\frac{\Gamma}{\omega
_0^2}(l+1)-(l-1)(l+1)(l+2)\biggr]a_l
\\
-\frac{2R^2}{\lambda^2}\biggl[\frac{l^2}{4l^2-1}\,
a_{l-1}+\frac{(l+1)(l+2)}{(2l+1)(2l+3)}\, a_{l+1}\biggr]=0 ,
\end{multline}
where
\begin{equation}\label{f6}
\omega _0^2=\frac{\alpha}{R^3}\,\frac{\rho}{(\varDelta\rho )^2}\, ,\;\;\;
\Gamma=\frac{\rho\rho '}{(\varDelta\rho )^2}\,\frac{1}{KR}\, ,
\end{equation}
and $a_l$ is the perturbation amplitude corresponding to $l$-th spherical harmonic
from a sum $\zeta (t)=\sum _la_l(t)Y_l$. Thus, we have a determinant of the
infinite order for determining the proper values. In the lack of gravity $\lambda
=\infty$ the off-diagonal terms vanish. The roots of the master equation give the
oscillation spectrum of a spherical solid with damping. All the frequencies have a
negative imaginary part, resulting in the conclusion that in this case the
spherically shaped solid is stable against small perturbations of its equilibrium
shape.
\par
In the presence of gravity we must involve the off-diagonal terms in (\ref{f5}).
Note that the diagonal terms increase proportional to $l^3$ as
$l\rightarrow\infty$, while the off-diagonal terms remain finite and have the
order of $(R/\lambda )^2$. This means that, as the degree of harmonic $l$
increases, the relative effect of the pressure gradient reduces, agreeing with the
stability of the flat interface at large wave vectors. As in the case of the flat
interface, the instability develops in the first turn for the minimum degree of
harmonics $l=2$.
\par
The spherical shape of helium crystals in the hcp phase is observed above the
first roughening transition temperature at $T>$1.25~K (Fig.~\ref{fig2}). On the
other hand, the hcp phase is limited by the hcp-bcc transition at $T=$1.44~K.
Within this temperature range the kinetic growth coefficient is small and varies
insignificantly with the temperature from $1/K\sim$11~m/s at the bcc transition to
$\sim$3~m/s near the roughening transition. Inserting these values into
Eqs.~(\ref{f5}, \ref{f6}) and solving numerically the master equation for proper
frequencies, we determine the critical radius $R_c=$0.61~cm. The radius of
crystals studied in \cite{Ts} does not exceed this magnitude. This may serve as an
explanation that all the crystals grown in that experiment display stability of
their spherical shape.
\par
\begin{figure}
\includegraphics[scale=0.6]{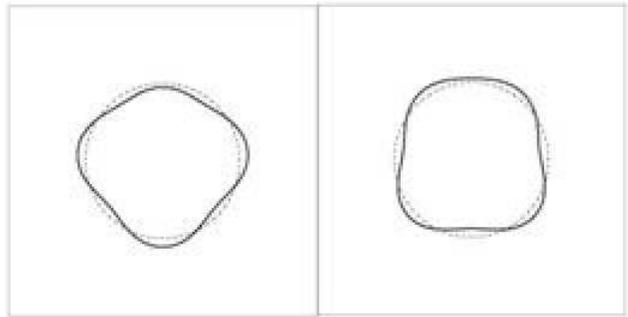}
\caption{The initial stage for the development of instability for a spherical
crystal with the size larger than the critical radius. The dashed circle is the
initial shape.} \label{fig3}
\end{figure}

The proper vectors in (\ref{f5}) are determined with accuracy to their sign. For
$R>R_c$, this gives two possible cases for developing the instability shown in
Fig.~\ref{fig3}.
In the first case at the bottom there appears a hogging in the upward direction.
In the second case we see a formation of constriction. Note that the result refers
only to the initial stage of developing instability.
\subsection{Crystals of the limited sizes}
\par
Another factor which stabilizes the flat surface is a finiteness of the crystal
size. In this case the spectrum of crystallization waves becomes discrete. The
instability will appear at the minimum wave vector. To estimate, we replace the
hexagonal crystal shape with the equivalent cylindrical one of radius $R$. Then we
can show that dispersion equation (\ref{f1}) holds its form but the perturbation
amplitude $\zeta$ of the flat crystal surface is given by
\begin{equation*}
\zeta(\bm{r},\, t)=\zeta _0J_m(qr)\, e^{im\phi}e^{-i\omega t},
\end{equation*}
where $\phi$ is the azimuth angle, $J_m$ is the Bessel function, and $r$ is the
distance from the center of a cylinder. For the circular symmetry at the fixed
volume of a crystal, the minimum wave vector $q_0$ is determined from the relation
\begin{equation}\label{f8}
J_1(q_0R)=0, \;\;\; q_0\approx 3.83/R\, .
\end{equation}
From expressions (\ref{f1}--\ref{f3}) and (\ref{f8}) one can obtain that the
instability should appear for a crystal with the diameter larger than 6~mm.
\section{Instability of a crystal facet}
\par
Below the roughening transition temperature a singularity appears as a nonanalytic
angular dependence in the behavior of the surface tension versus angle $\theta$
between the direction of crystallographic axis and the normal to the crystal
surface \cite{La}
\begin{equation*}
\alpha (\theta)=\alpha _0+\beta (T)\mid\theta\mid+\cdots , \;\;\; (|\theta |\ll
1).
\end{equation*}
Here $\beta$ is the  quantity proportional to the energy of a crystal facet step
and is positive below the roughening transition temperature. This entails that the
surface tension and surface stiffness represent the drastically different
quantities. The surface stiffness becomes infinite for zero angle $\theta =0$. The
singularity prevents from varying the shape of the crystal facet and qualitatively
changes the conditions for stability. Let us consider the distinctions in
appearing the Rayleigh instability in comparison with the case of the rough
crystal surface.
\subsection{Rayleigh instability of the horizontal crystal facet}
\par
Below the roughening transition the energy variation of the surface for
sufficiently small perturbation amplitude $\zeta (x,\, y)$ from the horizontal
flat facet is equal to
\begin{equation*}
\Delta E=\int\!\left(\frac{\alpha _0}{2}(\nabla\zeta
)^2+\beta\mid\nabla\zeta\mid-\,\frac{\varDelta\rho\, g}{2}\zeta ^2\right) dx\, dy
.
\end{equation*}
The smallness of $\zeta$ and $\nabla\zeta$ for the correct and consistent use of
the expansion of energy in $\zeta$ is provided by inequality $\beta\ll\alpha _0$.
Let us start from the one-dimensional case, namely, flat bar of length $L$. The
energy per unit length $\Delta E/D$ reads
\begin{equation*}
\frac{\Delta E}{D}=\int\limits _{-L/2}^{L/2}
\left(\frac{\alpha_0}{2}\biggl(\frac{\partial\zeta}{\partial
x}\biggr)^2+\beta\biggl|\frac{\partial\zeta}{\partial
x}\biggr|-\frac{\varDelta\rho\, g}{2}\,\zeta ^2\right) dx ,
\end{equation*}
where $D$ is the width of the crystal bar in the transverse direction. The
boundary conditions at the ends of the bar
\begin{equation}\label{f12}
\zeta (-L/2)=\zeta (L/2)=0
\end{equation}
correspond to the case when the crystal surface is immobile at these points. It is
convenient to introduce the dimensionless quantities according to $x'=x/L$, $\eta
=\zeta (\varDelta\rho\, gL)/\beta$, and $\gamma=\lambda /L$. Then,
\begin{equation*}
\frac{\Delta E[\eta (x')]}{D}=\frac{\beta ^2}{\varDelta\rho\, gL}\int
_{-1/2}^{1/2} \left(\frac{\gamma}{2}\,\eta ^{\prime\, 2} +\mid\eta
'\mid-\frac{\eta ^2 }{2}\right) dx'
\end{equation*}
The extremum satisfies the equation
\begin{equation*}
\gamma ^2\eta ^{\prime\prime}+2\delta(\eta ')\eta ^{\prime\prime}+\eta =0 .
\end{equation*}
Two types of functions can be solutions of the equation
\begin{equation*}
\eta (x')=\eta _0\left\{
\begin{array}{c}
1
\\
\sin(x'-x_0')/\gamma
\end{array}
\right. ,
\end{equation*}
i.e., either a constant or a sine function.
\par
The height of a potential barrier for destructing the flat crystal facet as well
as the type of a critical fluctuation depend on the magnitude $\gamma$. For
sufficiently small length of the facet $L<L_{c1}=\pi\lambda$, there exists a
single trivial solution $\eta (x')=0$, and the flat crystal facet remains stable.
For $L>L_{c1}$, there appears a nontrivial solution which consists of two
half-sinusoids and flat segment
\begin{equation*}
\eta (x')=\eta _0\left\{
\begin{array}{cc}
1\, , & \mid x'\mid\leqslant x_0'=\frac{1-\pi\gamma}{2}
\\
\\
\sin\frac{1/2-\mid x'\mid}{\gamma} \, , & x_0'<\mid x'\mid\leqslant 1/2
\end{array}
\right. .
\end{equation*}
Here the function $\eta (x')$ is continuous together with its derivative at points
$x'=\pm x_0'$ and vanishes at $x'=1/2$. The critical perturbation amplitude of
fluctuation is equal to
\begin{equation*}
\eta _{c1}=\frac{2}{1-\pi\gamma}\;\;\;\text{or}\;\;\;\zeta
_{c1}=\frac{2\beta}{\varDelta\rho\, g}\,\frac{1}{L-L_{c1}}\, .
\end{equation*}
The height of the potential barrier, separating the transition of the flat crystal
facet to a distorted state, is given by
\begin{equation}\label{f18}
\frac{\varDelta E}{D}=\frac{2}{1-\pi\gamma}=\frac{2\beta ^2}{\varDelta\rho\,
g}\,\frac{1}{L-L_{c1}}\, .
\end{equation}
As the size of a crystal facet increases, there appear additional possibilities
for other fluctuations consisting of a combination of flat segments and
half-sinusoids. The corresponding solutions, composed with $n$ flat segments,
appear as $L>L_{c\, n}$ where $L_{c\, n}=nL_{c1}$. As the number of the flat
segments increases, the perturbation amplitude of critical fluctuations does as
well
\begin{equation*}
\eta _{c\, n}=\frac{2n}{1-\pi\gamma\, n}\;\;\;\text{or}\;\;\;\zeta _{c\, n
}=\frac{2\beta}{\varDelta\rho\, g}\,\frac{n}{L-L_{c\, n}}\, .
\end{equation*}
The potential barrier height grows as the number $n$ of possible flat segments
increases
\begin{equation*}
\frac{\varDelta E_n}{D}=\frac{2n^2}{1-\pi\gamma\, n}=\frac{2\beta
^2}{\varDelta\rho\, g}\,\frac{n^2}{L-L_{c\, n}}\, .
\end{equation*}
Provided the experimental conditions correspond to the conservation of the total
mass including the both liquid and solid phases, the solutions should satisfy an
additional requirement
 $$
\int _{-L/2}^{L/2}\zeta (x)\, dx =0 .
 $$
In this case the nontrivial solutions for critical fluctuations can be realized
for the even numbers $n$ and, correspondingly, first critical length becomes equal
to $L_{c\, 2}$.
\par
For the temperatures well below the roughening transition, the facet step energy
coefficient $\beta$ is measured \cite{Bal}, and the numerical estimate of the
coefficient in Eq.~(\ref{f18}) gives the magnitude of the potential barrier about
10$^{-5}$~erg or 10$^{11}$~K for the basal facet of size $\sim$1~cm. The
overcoming of such barrier is practically impossible during the time of
experiment. Thus, the appearance of the singular angle dependence in the function
$\alpha (\theta )$ results in a drastic change for stability conditions of the
crystal surface. With very small cooling below the roughening transition
temperature the potential barrier becomes sufficiently large and the probability
of its overcoming due to thermal or quantum fluctuations is vanishingly small.
This conclusion is confirmed by the experimental evidence for stability of the
crystal shape below the roughening transition.
\par
The one-dimensional problem and boundary conditions (\ref{f12}) are chosen as
simplest ones in order to illustrate the method of solution and to obtain an
analytical estimate of the potential barrier height for developing gravitaional
Rayleigh-like instability. The full problem should be solved employing the real
crystal shape with the surfaces connecting the facets. We have analyzed an onset
of instability at the circular facet on the analogy with Sec.~IIC. The
mathematical treatment becomes more complicated but the final result for the
barrier height differs from Eq.~(\ref{f18}) with a numerical coefficient of about
unity.
\subsection{Instability of the tilted crystal facet}
\par
Here we analyze the stability of a crystal facet tilted with angle $(\pi
/2-\varphi )$  against its small perturbations. The variation of energy $\Delta E$
per width of the crystal facet reads
\begin{multline}\label{f21}
\frac{\Delta E}{D}=\int _{-L/2}^{L/2}dx\biggl(\frac{\alpha _0}{2}\,\zeta
^{\prime\, 2}(x)+\beta\mid\zeta '(x)\mid
\\
-\varDelta\rho\, g\big(x\zeta (x)\cos\varphi\, +\frac{\zeta
^2(x)}{2}\,\sin\varphi\bigr)\biggr) ,
\end{multline}
where $\zeta (x)$ is a perturbation amplitude taken from the flat surface. The
angle $\varphi =0$ means the vertical position of the crystal and $\varphi =\pi
/2$ corresponds to the horizontal position.
\par
Let axis $Oz$ run along the normal to the crystal facet. The axis $Oy$ is
perpendicular to the axis $Oz$ and the acceleration of gravity as well. The third
axis $Ox$, lying in the plane of the facet, is perpendicular to the axes $Oy$ and
$Oz$. The axis $Ox$ runs at the angle $\varphi$ to the direction of the
acceleration of gravity.
\par
Let perturbation amplitude of the crystal facet $\zeta (x)$ from its initial
position $\zeta (x)=0$ take place in the direction normal to the facet surface and
be independent of the coordinate $y$ directed horizontally along the crystal
facet. In addition, we assume that the small perturbation amplitude $\zeta (x)$ is
finite within the region $-L/2<x<L/2$ and the following boundary conditions are
fulfilled
\begin{equation}\label{f22}
\zeta (\pm L/2)=0\, .
\end{equation}
Here $L$ is implied as a size of the crystal shape fluctuation. The surface
fluctuations satisfy the conservation of the total mass of the liquid and solid
phases
\begin{equation}\label{f23}
\int _{-L/2}^{L/2}\zeta (x)\, dx=0\, .
\end{equation}
For $\varphi =\pi/2$, expression (\ref{f21}) goes over to the expression for a
horizontal crystal facet with the small vertical perturbations. For $\varphi =0$,
we have the expression for the energy of a vertical crystal facet with the small
perturbations in the direction normal to the facet.
\par
To find the minimum magnitude of the potential barrier which prevents from the
development of instability, we must consider extremum of functional (\ref{f21}).
The extremum of functional (\ref{f21}) satisfies the equation
\begin{equation}\label{f24}
\bigl(\alpha _0\zeta '+\beta\,\text{sgn}\,\zeta '\bigr)'+\varDelta\rho\,
g\bigl(x\cos\varphi +\zeta \sin\varphi\bigr)=0\, .
\end{equation}
Provided the derivative $\zeta '(x)$  does not vanish for all $x$ within
$-L/2<x<L/2$, i.e.,
\begin{equation}\label{f25}
\zeta '(x)\neq 0 ,
\end{equation}
equation (\ref{f24}) takes the simple form
\begin{equation}\label{f26}
\alpha _0\zeta ^{\prime\prime}(x)+\varDelta\rho\, g\bigl(x\cos\varphi +\zeta
(x)\sin\varphi\bigr)=0\, .
\end{equation}
If the condition (\ref{f25}) is satisfied, the general solution of Eq.~(\ref{f26})
reads
\begin{equation}\label{f27}
\zeta (x)= -x\cot\varphi +A\sin(x/\lambda _{\varphi})+B\cos(x/\lambda _{\varphi})
.
\end{equation}
Unknown coefficients $A$ and $B$ are determined by conditions (\ref{f22}) and
(\ref{f23}). Finally, we arrive at
\begin{equation}\label{f28}
\zeta (x)=\left(\frac{\sin (x/\lambda _{\varphi})}{\sin (L/2\lambda
_{\varphi})}-\,\frac{2x}{L}\right)\frac{L}{2}\cot\varphi\,  .
\end{equation}
Here $\lambda _{\varphi}$ plays role of an effective capillary length and
determines the typical scale of length at which $\zeta (x)$ varies
\begin{equation*}
\lambda _{\varphi}=\sqrt{\frac{\alpha _0}{\varDelta\rho\, g\sin\varphi}}
=\frac{\lambda}{\sqrt{\sin\varphi}}\, .
\end{equation*}
Note that, as $\varphi\rightarrow 0$ for the vertical position of a crystal facet,
the typical length $\lambda _{\varphi}$ diverges.
\par
Substituting (\ref{f28}) into (\ref{f21}), we have
\begin{multline}\label{f30}
\frac{\Delta E}{D}=\frac{\alpha _0}{2}\,\frac{\lambda _{\varphi}^3}{\lambda
^2}\,\frac{\cos ^2\varphi}{\sin\varphi}\int
_{-\tilde{L}/2}^{\tilde{L}/2}dx\biggl(a^2\cos 2x -2a\cos x
\\
+1+x^2+2(\beta /\alpha _0)\mid a\cos x -1\mid\tan\varphi\biggr)
\end{multline}
Here we have introduced the dimensionless quantities
 $$
\tilde{L}=\frac{L}{\lambda _{\varphi}}\;\;\;\text{and}\;\;\; a=
\frac{\tilde{L}/2}{\sin (\tilde{L}/2)}\, .
 $$
The magnitude of energy (\ref{f30}) can be written as
\begin{multline}\label{f31}
\frac{\Delta E}{D}=\frac{\alpha _0L}{4}\,\frac{\lambda _{\varphi}^2}{\lambda
^2}\,\frac{\cos
^2\varphi}{\sin\varphi}\biggl(\frac{\tilde{L}^2}{6}-2+\tilde{L}\cot\frac{\tilde{L}}{2}\biggr)
\\
+\beta\,\frac{\lambda _{\varphi}^3}{\lambda ^2}\,\cos\varphi\int
_{-\tilde{L}/2}^{\tilde{L}/2}dx \mid a\cos x -1\mid \, .
\end{multline}
For $\tilde{L}=L/\lambda _{\varphi}\ll 1$, equation (\ref{f31}) gives
\begin{equation*}
\frac{\Delta E}{D}=-\,\frac{\alpha _0L}{80}\,\cot ^2\varphi\left(\frac{L}{2\lambda
_{\varphi}}\right) ^4 +\frac{\beta L}{12}\cot\varphi\left(\frac{L}{2\lambda
_{\varphi}}\right) ^2 .
\end{equation*}
As a result, for sufficiently small-sized fluctuations with $L\ll L_{cr}$, the
energy of fluctuations is positive $\Delta E>0$ and such fluctuations are of low
probability. The tilted facet is practically stable for such perturbations. The
critical length is equal to
\begin{equation*}
L_{cr}=4\,\lambda _{\varphi}\sqrt{\frac{5\beta}{3\alpha _0}\,\tan\varphi}=\frac{4
\sqrt{5\beta /(3\varDelta\rho\, g)}}{\sqrt{\cos\varphi}}\, .
\end{equation*}
This means that the crystal facet is practically stable if its size does not
exceed the critical length. In fact, the magnitude of the energy barrier $\Delta
E\sim\beta ^2/\varDelta\rho\, g$ proves to be about 10$^{-5}$~erg or 10$^{11}$~K.
As the size of a crystal becomes larger than the critical length $L>L_{cr}$, the
tilted crystal facet becomes unstable against distortion of its shape.
\par
The analysis of the crystal facet stability for the larger lengths $L\sim\lambda
_{\varphi}$ is simple. Let us represent Eq.~(\ref{f31}) in the form of the
following inequality
\begin{multline*}
\frac{\Delta E}{D}\leqslant\frac{\alpha _0L}{4}\,\frac{\lambda
_{\varphi}^2}{\lambda ^2}\,\frac{\cos
^2\varphi}{\sin\varphi}\biggl(\frac{\tilde{L}^2}{6}-2+\tilde{L}\cot\frac{\tilde{L}}{2}\biggr)
\\
+\beta\,\frac{\lambda _{\varphi}^3}{\lambda ^2}\,\cos\varphi\biggl(\int
_{-\tilde{L}/2}^{\tilde{L}/2}\! dx\, (\mid a\cos x \mid +1 )\biggr) .
\end{multline*}
The right-hand side of the inequality is obviously negative for
$\tilde{L}\gtrsim\pi$, i.e. $\Delta E <0$. This means that a crystal facet
distortion with such lengths is energetically favorable. The tilted crystal facet
becomes absolutely unstable if the facet size $L\gtrsim\lambda _{\varphi}$.
\par
\begin{figure}
\vspace*{1em}
\includegraphics[scale=0.6]{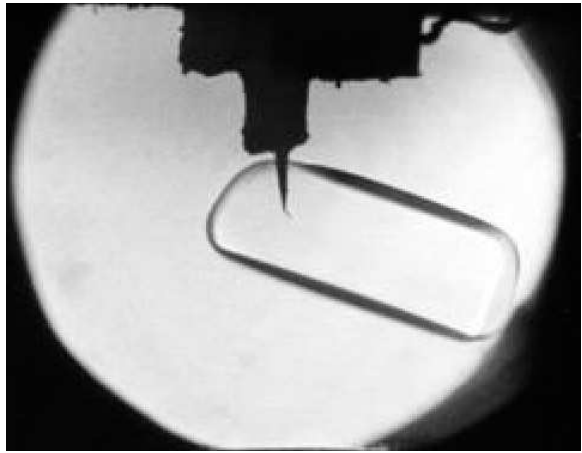}
\caption{The lateral facet of a $^4$He crystal is stable. The temperature is below
the second roughening transition by about 20~mK.} \label{fig4}
\end{figure}

In conclusion, such high potential barriers can explain a gravitational stability
of a crystal facet in the immediate vicinity of the roughening transition
temperature at which the facet step energy vanishes (Fig.~\ref{fig4}).
In Fig.~\ref{fig4} the image of a $^4$He crystal at 0.901~K is given. Note that
the lateral faceting disappears at 0.910~K, demonstrating the roughening
transition for the $a$-facets. The crystal has a clear lateral facet which
remained stable in the course of experiment during, at least, 10~min.
\section{Conclusion}
\par
The gravitational instability at the atomically rough $^4$He surface, which
develops under the lack of any potential barrier, is similar to the classical
Rayleigh instability  when the heavier liquid lies above the lighter liquid. The
distinction is that the time necessary for the development of instability is
determined by the kinetic growth coefficient of a crystal surface.
\par
As for the smooth faceted crystal surface, having a singularity in the surface
stiffness, the development of the surface  instability becomes possible due to
thermal or quantum fluctuations if the size of the facet surface exceeds the
critical one. However, the large height of a potential barrier makes its
overcoming impossible for a experimentally reasonable time. This explains the
stability of the lower facet for a free-growing $^4$He crystal observed in the
experiment during a few hours.

\end{document}